\documentclass{aa}
\usepackage{psfig,latexsym,longtable}

\begin{document} 

\thesaurus{03(13.25.2;11.19.1;12.03.3)}

\title{A Hard Medium Survey with ASCA.III.: a Type 2 AGN revealed from X-ray spectroscopy
\thanks{Based on observations performed at the European Southern Observatory, Paranal, Chile}}

\author{R. Della Ceca \inst{1}, T. Maccacaro \inst{1}, P. Rosati \inst{2} and V. Braito \inst{1}}

\institute{
Osservatorio Astronomico di Brera, via Brera 28, I-20121 Milano, Italy.
\and
European Southern Observatory, D-85748, Garching bei Munchen, Germany.
}

\date{Received: November 12, 1999; Accepted: December 20, 1999}
\authorrunning{Della Ceca et al.,}
\titlerunning{A Hard Medium Survey with the ASCA GIS.III}
\maketitle

\begin{abstract}

In this paper we report the discovery of an hard X-ray selected Type 2
Seyfert galaxy and we present and discuss its X-ray and optical spectrum 
together with the radio to X-ray energy distribution. The X-ray source - AXJ2254+1146
- is part of the ASCA Hard Serendipitous Survey (HSS).
What makes this discovery particularly noteworthy is the fact that the 
Type 2 classification of this Seyfert galaxy has resulted directly from the 
X-ray data and has been confirmed by optical spectroscopy only subsequently.

The X-ray spectrum of AXJ2254+1146 is best described by a model consisting 
of an unresolved Gaussian line at $6.43\pm 0.1$ keV 
plus the so called ``leaky-absorber" continua having an intrinsic 
power law photon index of $\Gamma$ = $2.51^{2.76}_{2.17}$
(1 $\sigma$ confidence interval).
The best fit values of the absorbing column density ($N_H$ =
$1.85^{2.24}_{1.47} \times 10^{23}$ cm$^{-2}$), of the line equivalent
width ($0.6^{0.84}_{0.36}$ keV) and of the scattering fraction
($0.7^{1.4}_{0.1} \%$),  lead us to classify it as a Type 2 AGN from an 
X-ray point of view.

Inspection of the POSS II image reveals the presence, within the ASCA
X-ray error circle, of the nearby Sbc spiral galaxy UGC 12237 
($m_{B_o}=14.26$)
that, even
on positional ground considerations alone, is the most likely optical
counterpart of AXJ2254+1146. 
Subsequent optical spectroscopy of UGC 12237 has confirmed its Seyfert 2 
optical nature.

\end{abstract}

\section{Introduction}


Absorbed (Type 2) AGN have been proposed as the major contributors to
the Cosmic X-ray Background (CXB)
above 2 keV (see among others Setti and Woltjer, 1989; Madau, Ghisellini and
Fabian, 1994; Comastri et al., 1995).  Indeed, recent results from ASCA
and Beppo-SAX observations (Akiyama et al., 1998a;
Bassani et al.,1999; Risaliti, Maiolino and 
Salvati, 1999; Fiore et al., 1999) seem to favor this hypothesis  but deeper
investigations are still needed to confirm and refine this scenario and
to test competing models.  
For example it is not clear if high
luminosity Type 2 AGN exist or not (see e.g. Halpern, Turner and George, 1999;
Akiyama et al., 1998a) or if a significant fraction of the hard CXB
deriving from star-burst and star-forming galaxies at moderate redshift
can be excluded or not (see e.g.  Moran, Lehnert and Helfand, 1999).

At the {\it Osservatorio Astronomico di Brera}, with the specific aim of 
extending to faint fluxes the census of the X-ray sources shining in 
the hard X-ray sky, we have initiated a few years ago the ASCA Hard 
Serendipitous Survey (HSS: see Della Ceca at al., 1999a for a progress 
update): a systematic search for sources in the $2-10$ keV 
band, using data from the GIS2 instrument on board the ASCA satellite 
(Tanaka, Inoue and Holt, 1994). 
This effort has lead to a ``pilot" sample of 60 sources (Cagnoni, Della Ceca and 
Maccacaro, 1998) whose spectral properties 
have been presented and discussed by Della Ceca at al., (1999b).
The ASCA HSS has been extended and it now covers $\sim 71$ deg$^2$ of sky.
The resulting source sample consists of 189 sources detected with a 
signal-to-noise ratio $\geq 4.0$ (Della Ceca at al., 1999a, 2000), a
more restrictive criterion than that used in Cagnoni, Della Ceca and Maccacaro, 
1998 where a signal-to-noise ratio $\geq 3.5$ was used.

In this paper we report the serendipitous discovery of a low luminosity
Type 2 AGN: AXJ2254+1146. 
Its absorbed Type 2 nature has been revealed by
the X-ray data alone, thanks to the good spectral resolution of the ASCA
GIS ($ \Delta E/E \sim 8\%$ at $6-7$ keV).
We are convinced that the new generation of X-ray telescopes 
(e.g. Chandra, XMM) will make
AXJ2254+1146 only the first one of several 
objects  whose nature will be discovered by means of the X-ray data alone.

The paper is organized as follows.  In section 2 we present the
X-ray data and discuss the X-ray spectral properties of AXJ2254+1146.  
In section 3 we present its optical identification (UGC 12237), while 
in section 4 we discuss the overall
Spectral Energy Distribution (SED) of the source and compare it with
the SED of Seyfert 2 and non-AGN spiral galaxies.
An optical spectrum of UGC 12237 is shown and discussed in section 5.
Finally,
summary and conclusions are presented in section 6.  We adopt $H_{0} =
50$ km s$^{-1}$ Mpc$^{-1}$ and $q_{0} = 0$ throughout.

\section {X-ray Data and Analysis}

\subsection {X-ray Imaging}

AXJ2254+1146 was discovered in the ASCA GIS field 74076000 
\footnote{The ASCA satellite also has two solid-state imaging
spectrometer (SIS) on the focal plane. Unfortunately this observation
was operated in the ``2 CCD mode"  configuration and AXJ2254+1146 is
out of the investigated sky region. So we do not have SIS data on this source.}
pointed at
the radio loud quasar PKS2251+113 ($\alpha_{2000}=$
$22^{h}54^{m}10.5^{s}$ and $\delta_{2000}= +11^{o}36'38"$; z=0.323).
The field is one of the many used to extend the survey of Cagnoni, Della Ceca and
Maccacaro 1998; data preparation and analysis have been discussed in detail therein and in
Della Ceca et al., 1999b.
This source attracted our attention because its hardness ratios 
(HR1 = $0.41\pm 0.12$; HR2 = $0.47\pm 0.06$)
\footnote{
HR1 and HR2 are defined in the following way: 
$$HR1 = {M-S \over M+S}\ \ \ \ \ \ \  HR2 = {H-M \over H+M}$$
where S, M and H are the corrected net counts in the 0.7-2.0, 2.0-4.0
and 4.0-10.0 keV energy band respectively (cf. Della Ceca et al., 1999b).
         }
can not be explained by a single absorbed power law model, implying a 
more complex and structured X-ray spectrum (cf. Della Ceca et
al., 1999b).

Panel (a) of Figure 1 shows  the central ($32' \times 32'$) smoothed
\footnote {The smoothing function is a two-dimensional Gaussian
filter with $\sigma = 1'$, which is comparable to the ``core" of the
(XRT+GIS2) point-spread function (PSF).}
GIS2 hard (2-10 keV) image with overlaid a contour plot indicating the
X-ray emission intensity.
Three sources are clearly visible in the 2-10 
keV band image: the target of the observation, (AXJ2254+1136 $\equiv$ 
PKS2251+113), plus two serendipitous sources, AXJ2254+1141 and 
AXJ2254+1146. AXJ2254+1141 is identified with MS2252.2+1126, a Type 1 
AGN at z=0.0281 (Stocke et al., 1991, Maccacaro et al, 1994).  

Panel (b) shows a ``hard-over-soft" image obtained dividing the smoothed
``hard" (2-10 keV) image by the smoothed ``soft" (0.7 - 2.0 keV)
image. The same contours of panel (a) have been overlaid on the image for
ease of comparison. AXJ2254+1146 stands out as a very hard source compared
to the other ones.

  
In Table 1 the X-ray position, the optical position and the offset
between the two (X-ray - optical) are listed for the three sources
previously mentioned. The measured offsets are consistent
with the positional uncertainties of the ASCA GIS images
\footnote {The positional uncertainty of the ASCA detected sources has been
investigated by Gotthelf (1996) using 48 point-like sources. He reached the
conclusion that, in the case of the combination XRT+SIS, the 90\%
confidence level error circle is of the order of $\sim 40"$ radius, while in
the case of the XRT+GIS additional instrumental and calibration
uncertainties can produce larger errors. From the analysis of a sample of 168 
point-like ASCA targets within the data set used for the ASCA HSS we 
have estimated a 90\% confidence level error circle of $\sim 2$
arcmin radius.}
.

\begin{table*}
\begin{center}
\caption{X-ray Sources in the ASCA GIS2 field 74076000}
\begin{tabular}{lccccc}
\hline
\hline
Name         & Ra; Dec                 & Ra; Dec                & Offset                     & ID          & Note  \\
             & (J2000)                 & (J2000)                & $\Delta$ RA; $\Delta$ Dec  &             &       \\
             & X                       & Optical                & (arcsec)                   &             &       \\
\hline
\hline
AXJ2254+1136 & 22 54 11.3; +11 35 51.5 & 22 54 10.5; +11 36 38 & $-$11.8; +46.5   & PKS2251+113   & RL Type 1 AGN; target                \\
AXJ2254+1146 & 22 54 20.5; +11 46 05.7 & 22 54 19.7; +11 46 57 & $-$11.8; +51.3   & UGC 12237     & Type 2 AGN; this paper \\
AXJ2254+1141 & 22 54 43.4; +11 41 51.2 & 22 54 43.7; +11 42 49 & +4.4; +57.9    & MS2252.2+1126 & Type 1 AGN            \\

\hline
\hline
\end{tabular}
\end{center}
\end{table*}

\begin{figure}
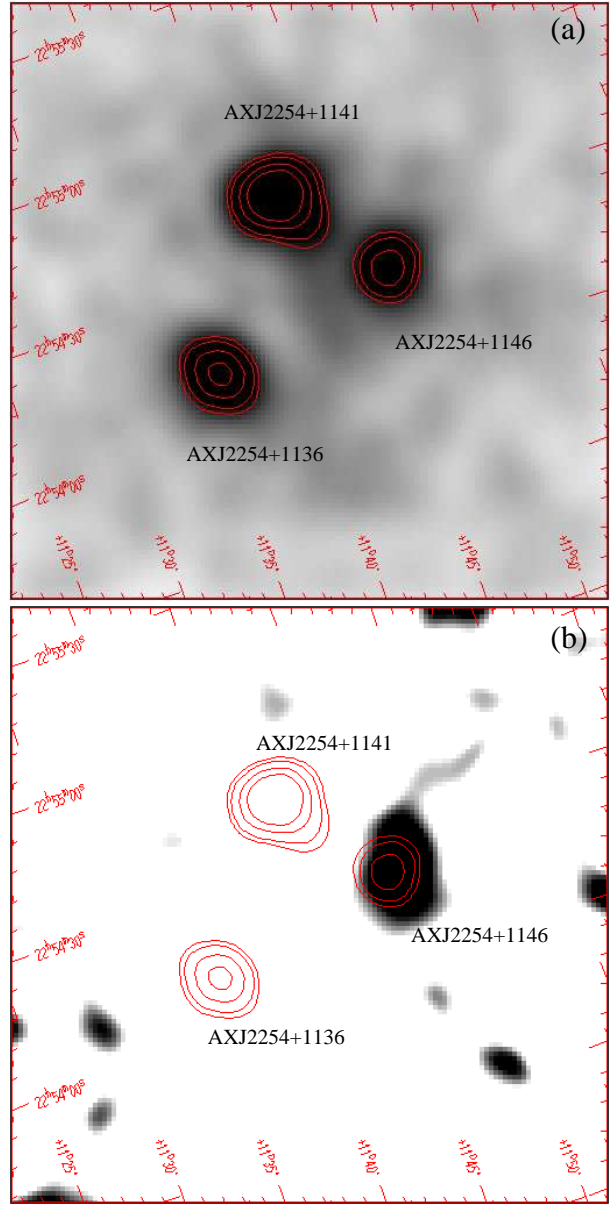

\psfig{file=AA9395.f1a,height=8.0cm,width=8.0cm}
\psfig{file=AA9395.f1b,height=8.0cm,width=8.0cm}
\caption{Panel a: 2-10 keV image of the central part 
($32' \times 32'$) of
the GIS2 field of view. The raw data have been
smoothed with a two-dimensional Gaussian filter with $\sigma = 1'$;
contours are at 2.5, 3, 4, 5 and 10 sigma above the background.  
Panel b: the same contours have been overlaid on the ``hard-divided-by-soft" 
image of the field (see text for details).
Note the non-standard orientation of the axis.}
\end{figure}

\subsection {X-ray Spectrum}

To maximize statistics and signal-to-noise ratio, total counts
(source + background) were extracted from a circular region of 2 arcmin
radius  around the source centroid in the GIS2 and GIS3 images.
Background counts  were taken from two circular uncontaminated regions
of 4.75 arcmin radius close to the source.  Source and background data
were extracted in the ``Pulse Invariant" (PI) energy channels, which
have been corrected for spatial and temporal variations of the detector
gain.  The Ancillary Response File (ARF) was created with version 2.72
of the FTOOLS task ASCAARF at the location of the source in the
detector.  In order to improve the statistics, we have produced a
combined GIS spectrum (adding the GIS2 and GIS3 data) and the
corresponding background and response matrix files, following the
recipe given in the ASCA Data Reduction Guide (rev 2.0; see section
8.9.2 and 8.9.3 and reference therein) and using the FTOOLS 4.2
software package (supplied by HEASARC at the Goddard Space Flight Center).
The total ($\equiv$ GIS2 + GIS3) net counts in the source extraction
region and in the energy interval used (0.7 -- 10 keV) are $290 \pm
20$, overimposed over a total $\sim 110$ background counts.
In order to use the $\chi^{2}$ statistics in the spectral
fitting procedure the combined GIS spectrum was rebinned to give at
least 15 total counts per energy bin.  Spectral analysis has been
performed using XSPEC 10.0.  All the models discussed below have been
filtered by the Galactic absorption column density along the line of
sight ($N_{H_{Gal}} = 5.2\times 10^{20}$ cm$^{-2}$; Dickey and Lockman, 1990).
All the errors reported in the spectral analysis are at the 68\%
confidence level for 1 interesting parameter ($\Delta \chi^{2} =
1.0$), unless explicitly quoted.

Single-component models do not provide an adequate description of the
spectrum of AXJ2254+1146. A single temperature Raymond-Smith model or 
a single power-law model are both rejected at a high confidence level ($\geq
99.999\%$).

We have first tried a single absorbed power law model.  
The best fit values for the power law photon index ($\Gamma$) and
for the absorbing column density ($N_H$) are
$2.18^{2.80}_{1.68}$ and $1.67^{2.15}_{1.32} \times 10^{23}$ cm$^{-2}$, 
respectively.  
Although this fit is formally acceptable ($\chi^{2} = 29.6$ for 23
degree of freedom) the residuals (reported in figure 2) show an excess
below $\sim$ 2 keV and also around $\sim$ 6.5 keV;  
a more complex spectrum is thus needed to describe the overall spectral
properties of AXJ2254+1146.

\begin{figure}
\psfig{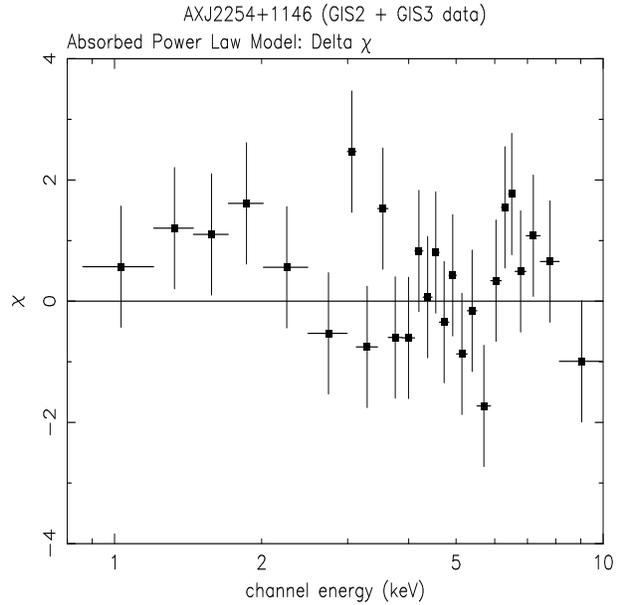}
\caption{Residuals of the 
combined GIS2+GIS3 data set compared with the best fit absorbed power law model.}
\end{figure}

We have then tried a spectral model consisting of a narrow (unresolved)
Gaussian line at $\sim$ 6.4 keV, plus the so
called ``leaky-absorber" continua; the latter being 
composed by an absorbed plus a non-absorbed power law model having the
same photon index.  The absorbed power law model represents the ``first
order" AGN spectrum transmitted through an absorbing cold medium (torus
?) while the non-absorbed power law model represents the primary AGN
spectrum scattered into our line of sight by a warm, highly ionized gas
located outside the absorbing medium.  The scattered fraction is given
by the ratio between the normalization of the non-absorbed and absorbed
power law components.

The results are reported in Table 2, while in Figure 3 we show the
unfolded and folded spectrum and the residuals to the best fit
model. 
The best fit values are 
$\Gamma$ = $2.51^{2.76}_{2.17}$ and $N_H$ = $1.85^{2.24}_{1.47}
\times 10^{23}$ cm$^{-2}$.  
The scattering fraction implied from the
fit is $0.7^{1.4}_{0.1} \%$; the position and the observed equivalent width 
(EW$_{obs.}$ hereafter) of
the line are $6.43^{6.52}_{6.33}$ keV and $0.6^{0.84}_{0.36}$ keV,
respectively.  
Note that EW$_{obs.}$ has been computed against the 
observed total (i.e. absorbed plus non-absorbed) source continuum.
As evident from the residuals shown in figure 2  and from the
$\chi^{2}_{\nu}$ in Table 2 this model represent a good description of
the spectrum of AXJ2254+1146.
We have also tried a fit with the width of the line as a free
parameter; the other best fit values are unaffected from this new fit,
while, at the spectral resolution of the GIS2 instrument, the line
is unresolved with a $1\sigma $ upper limit on the width of 0.6 keV.
The best fit intrinsic photon index ($\Gamma$ =
$2.51^{2.76}_{2.17}$) is steeper than the mean observed photon index in
Seyfert 1 galaxies ($\Gamma = 1.7\pm 0.1$, Nandra et al., 1997). We
have then tried to fit the data using the ``leaky absorber" continua
plus the narrow Gaussian line model but fixing $\Gamma = 1.7$; the
absorbing column density, the scattered fraction, the  EW$_{obs.}$ and
the position of the line are consistent, within the errors, with that
reported in table 2.

\begin{table*}
\begin{center}
\caption{Results of the Spectral Fit (GIS2 + GIS3 data): ``Leaky-absorber" continua + Narrow Gaussian Line.}
\begin{tabular}{cccccccc}
\hline
\hline
$\Gamma    $         &  Norm$_{abs.}$                     &  Norm$_{non-abs.}$                         & $N_H$    
                     &  E$_{Line}$                        &  Norm$_{Line}$                           
                     &  EW$_{obs.}$                       &  $\chi^{2}_{\nu}$/dof  \\
 (1)                 &  (2)                               &  (3)                                     &  (4)     
                     &  (5)                               &  (6)                                   
                     &  (7)                               &  (8)                    \\
\hline
\hline

$2.51^{2.76}_{2.17}$ & $2.51^{7.60}_{1.93}\times 10^{-3}$ & $1.70^{2.64}_{0.87}\times 10^{-5}$ & $1.85^{2.24}_{1.47}\times 10^{23}$
                     & $6.43^{6.52}_{6.33}$     & $1.07^{1.50}_{0.64} \times 10^{-5}$    
                     & $0.6^{0.84}_{0.36}$      & 0.94/20                \\

\hline
\hline
\end{tabular}
\end{center}
NOTE -- Allowed ranges are at 68\% confidence level for one interesting parameter ($\Delta \chi^{2} = 1.0$).
Columns are as follows: 
(1) Power-law photon index;
(2) Normalization of the absorbed component in units of 
    photons s$^{-1}$ cm$^{-2}$ keV$^{-1}$ at 1 keV; 
(3) Normalization of the non-absorbed component in units of 
    photons s$^{-1}$ cm$^{-2}$ keV$^{-1}$ at 1 keV; 
(4) Intrinsic absorbing column density in units of cm$^{-2}$;
(5) Line energy in keV;
(6) Line normalization. This number represents the total 
    photons cm$^{-2}$ s$^{-1}$ in the line;
(7) Observed equivalent width, i.e. the EW computed against the 
    observed total (i.e. absorbed plus non-absorbed) source continuum;
(8) Reduced $\chi^{2}$ and degree of freedom.
\end{table*}

\begin{figure}
\hskip -1truecm
\psfig{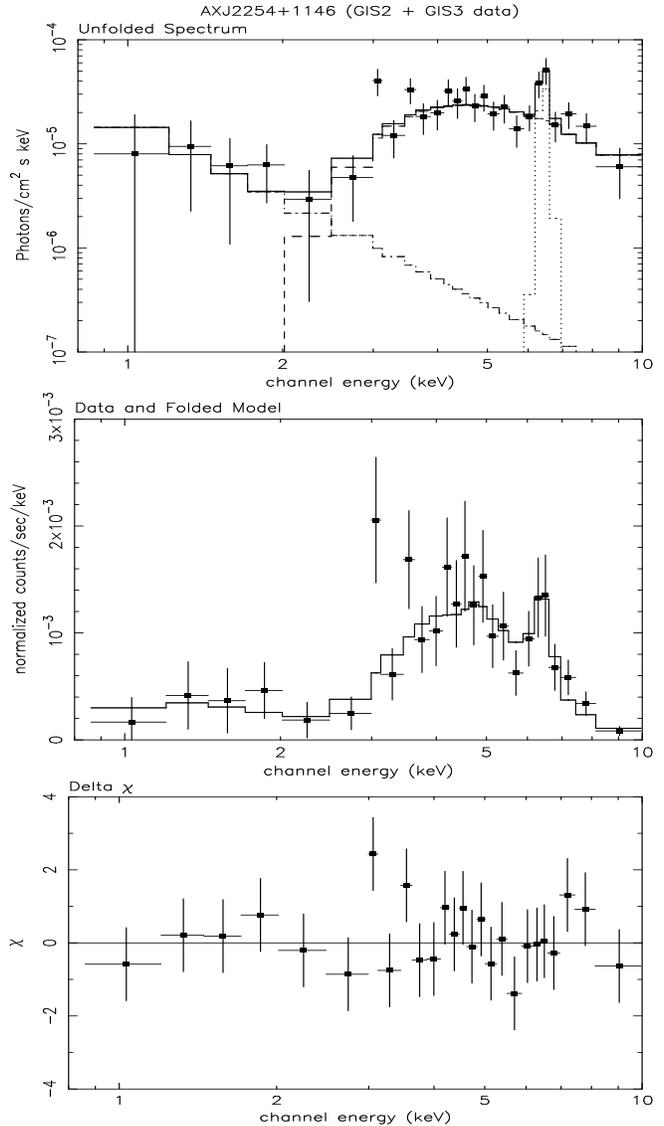}
\caption{Unfolded spectrum, folded spectrum and the residuals of the 
combined GIS2+GIS3 data set compared with the best fit model 
(narrow Gaussian line plus ``leaky-absorber" continua).}
\end{figure}

Seyfert 1 and Seyfert 2 galaxies are usually characterized by an
emission line at 6.4 keV produced by the K$_{\alpha}$ emission of Iron
species less ionized than Fe XVI, while in some Seyfert 2 object
emission lines from He-like (6.7 keV) and H-like Iron (6.96 keV) have
also been observed (see Turner et al., 1997a,b and reference therein).
The last two are emitted from highly ionized gas and are supposed to
derive from the warm scattering medium located outside the absorbing
material, i.e.  the same medium which is supposed to scatter the primary
AGN spectrum into our line of sight.
As shown in Matt et  al., 1996 the EW of the He-like and H-like Iron
lines, if computed with respect to the scattered component
(EW$_{scatt.}$ hereafter), can be as high as a few keV; these lines should
emerge in Compton thick system ($N_H > few \times 10^{24}$ cm$^{-2}$)
while they should be strongly diluted if the primary radiation is
directly visible (as in the case discussed here above 4 keV).
If we take an implausible upper limit of
EW$_{scatt.}$ = 10 keV  for the He-like or the H-like Iron line, we
should observe, in the case of AXJ2254+1146, a line with EW$_{obs}$
$\leq$ 0.07 keV; this is about a factor 9 less than that observed.
We can then rule out an origin of the observed line
from He-like and H-like iron species, leaving the fluorescent line at
6.4 keV produced by low ionization Iron the more plausible
possibility.
The measured position of the line
($6.43$ keV) is consistent, within its 95\% confidence range of 
($6.20 - 6.65$ keV), with the low ionization Iron $K_{\alpha}$ line 
emitted at the redshift of the object (z = 0.029, see section 5). 

The EW$_{obs.}$ is significantly larger than that observed in Seyfert 1
galaxies (EW = $230 \pm 60$ eV, Nandra et al., 1997) and is similar to
that measured in others Seyfert 2 galaxies (Bassani et al., 1999).
The measured EW ($\sim 0.6$ keV)
is about a factor 4.5 greater than that expected
from transmission through a spherical uniform shell of neutral material
(solar abundances) having the measured intrinsic $N_{H} \sim 2\times
10^{23}$ cm$^{-2}$ (Leahy and Creighton, 1993) and about 
a factor 7.5 greater than that expected to be produced from transmission
through an absorbing torus with a geometrical configuration like that
assumed by Ghisellini, Hardt and Matt, 1994 and having $N_{H} \sim
2\times 10^{23}$ cm$^{-2}$.  This could suggest a combined origin
either from the accretion disk and from the absorbing torus.
The scattering fraction ($\sim 1\%$) is consistent with that observed 
in other well studied Seyfert 2 galaxies (see Turner et al., 1997a,b and 
reference therein).

The Iron line equivalent width along with the scattering
fraction and the measured absorbing column density clearly allow us to
classify AXJ2254+1146 as a Type 2 object directly from X-ray
spectroscopy.  

We note that 
two data points at about 3 and 3.5 keV (see figure 3) seem to
be in excess with respect to the best fit model;
these line energies are in very good agreement
with the $K_{\alpha}$ neutral emission lines of Ar (2.96 keV) and Ca
(3.7 keV) respectively. Both lines are expected to be present in a
Compton reflected spectrum (see e.g the theoretical reflected X-ray
spectrum presented in Reynolds et al., 1994).

The observed (de-absorbed from the Galactic $N_H$ value) flux 
implied from the best fit model (see Table 2) in the 
0.5$-$2.0 keV, 0.5$-$4.5 keV and 2$-$10 keV energy band are 
$3.86\times 10^{-14}$ erg cm$^{-2}$ s$^{-1}$,  
$2.46\times 10^{-13}$ erg cm$^{-2}$ s$^{-1}$ and
$1.14\times 10^{-12}$ erg cm$^{-2}$ s$^{-1}$, respectively. 
The intrinsic luminosity (i.e. the luminosity emitted from the 
nucleus) in the 0.5$-$2.0 keV and in the 2$-$10 keV energy 
band are 
$2.35\times 10^{43}$ erg s$^{-1}$ and 
$1.30\times 10^{43}$ erg s$^{-1}$, respectively.

\section {Source Identification}

Figure 4 shows the Palomar Observatory Sky Survey II (POSS II) image
centered on the X-ray position of AXJ2254+1146 with its 90\% error
circle of 2.0 arcmin radius (see footnote 4).  We have also marked with
a cross the X-ray position of  AXJ2254+1146 if we apply the offset we
are measuring for the target of the ASCA observation (see table 1).

\begin{figure}
\psfig{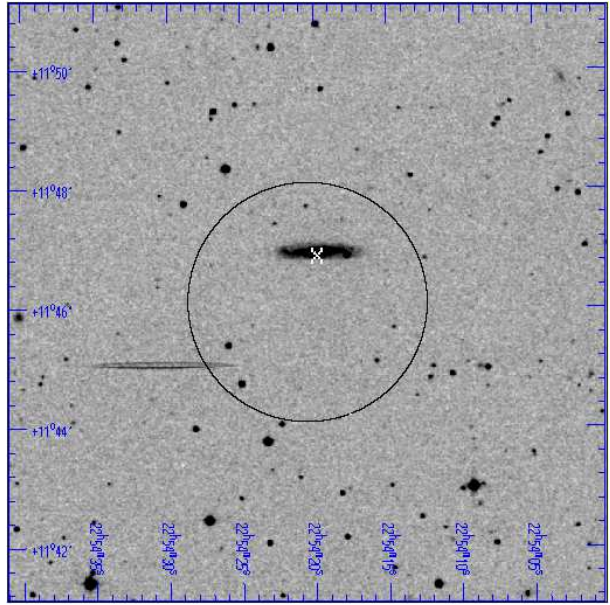}
\caption{POSS II image of the field centered on
AXJ2254+1146. The circle represents the  90\% confidence level error
box of 2 arcmin radius, while the white cross marks the position of 
the X-ray centroid when the offset measured for the target of the observation is
applied (see table 1).}
\end{figure}

The fairly bright ($m_{B_{o}} = 14.26$) 
spiral (Sbc) galaxy present in the error circle is 
UGC 12237.
Using the surface density of similarly bright galaxies of $<1$
deg$^{-2}$ (Maddox et al., 1990) we estimated that the probability for
a galaxy brighter than m = 15 to fall by chance in an X-ray error
circle of 2 arcmin radius is $<3.5 \times 10^{-3}$. This,
combined with the other data discussed in this paper (i.e. 
SED and the optical spectral properties), 
clearly shows that UGC 12237 is the most
probable optical counterpart of AXJ2254+1146.

\section {Spectral Energy Distribution}

UGC 12237 has been detected in the NVSS (Condon et al., 1998) 
as a 1.4 GHz
compact radio source 
and also in the IRAS survey at 25, 60 and
100 $\mu m$ (but not at 12  $\mu m$).  
There are no pointed ROSAT observations containing UGC 12237 nor is the galaxy
detected in the ROSAT All Sky Survey.  
The only available X-ray
measurement  is an upper limit of $\sim 2.5 \times
10^{-13}$ erg cm$^{-2}$ s$^{-1}$ in the 0.5-4.5 keV energy band from
the {\it Einstein} satellite (Burstein et al., 1997). 
This value is consistent with the 0.5 - 4.5 keV flux measured
here. Photometric data on this galaxy are reported in Table 3.

\begin{table*}
\begin{center}
\caption{Photometry of AXJ2254+1146 $\equiv$ UGC 12237}
\begin{tabular}{lcrlrr}
\hline
\hline
                    &  Band        &  Frequency            &  Observed Flux Density  &  $\nu$ $f_{\nu}$       &   $\nu$ $L_{\nu}$ \\
                    &              &      Hz               &                         & erg cm$^{-2}$ s$^{-1}$ & erg s$^{-1}$      \\
\hline
\hline
Radio (NVSS)        &  1.4 GHz      &  $1.4\times 10^{9}$  &  4.5 mJy        &  $6.3 \times 10^{-17}$   &  $2.5 \times 10^{38}$ \\
Infrared (IRAS)     & 100 $\mu$ m   &  $3.0\times 10^{12}$ &  $ 1.772$ Jy    &  $5.3 \times 10^{-11}$   &  $2.1 \times 10^{44}$ \\
Infrared (IRAS)     &  60 $\mu$ m   &  $5.0\times 10^{12}$ &  $ 0.642$ Jy    &  $3.2  \times 10^{-11}$  &  $1.3 \times 10^{44}$ \\
Infrared (IRAS)     &  25 $\mu$ m   &  $1.2\times 10^{13}$ &  $ 0.198$ Jy    &  $2.4 \times 10^{-11}$   &  $9.6 \times 10^{43}$ \\
Infrared (IRAS)     &  12 $\mu$ m   &  $2.5\times 10^{13}$ &  $<0.109$ Jy    &  $<2.7 \times 10^{-11}$  &  $<1.1 \times 10^{44}$ \\
Optical             &  B$^o$        &  $6.8\times 10^{14}$ &  14.26 mag      &  $5.8 \times 10^{-11}$   &  $2.3 \times 10^{44}$ \\
X-ray (ASCA)        &  1 keV        &  $2.4\times 10^{17}$ &  $2.0\times 10^{-14}$  erg cm$^{-2}$ s$^{-1}$ keV$^{-1}$      
                                                                             &  $2.0 \times 10^{-14}$   &  $8.0 \times 10^{40}$ \\
X-ray (ASCA)        &  5 keV        &  $1.2\times 10^{18}$ &  $1.9\times 10^{-13}$  erg cm$^{-2}$ s$^{-1}$ keV$^{-1}$      
                                                                             &  $9.3 \times 10^{-13}$  &   $3.7 \times 10^{42}$ \\
X-ray (ASCA)        &  10 keV       &  $2.4\times 10^{18}$ &  $1.0\times 10^{-13}$  erg cm$^{-2}$ s$^{-1}$ keV$^{-1}$      
                                                                             &  $1.0 \times 10^{-12}$  &   $4.0 \times 10^{42}$ \\
\hline
\hline
\end{tabular}
\end{center}
\end{table*}

In Figure 4 we show the Spectral Energy Distribution (SED) of
UGC12237:  the filled dots represent the photometric data from radio to
optical wavelengths while the tick solid line represents the $1 - 10$
keV best fit spectral model reported in Table 2.  Also shown in Figure
5 is the mean SED of a sample of Seyfert 2 and a sample of non-AGN
Spiral galaxies as determined by Schmitt et al., 1997. The SEDs
have been normalized at $\sim$ 2 keV.

It is clear from Figure 4 that the mean SED of non-AGN Spiral galaxies 
is at variance with the SED of  AXJ2254+1146, while 
the SED of Seyfert 2 galaxies seem to be more consistent with the data.
The largest discrepancy between the Seyfert 2 SED and the SED 
of AXJ2254+1146 is in the optical domain, where only the total magnitude 
(nucleus + host galaxy) is available; this discrepancy is expected 
since the SEDs reported in Schmitt et al., 1997 were based on the 
optical nuclear flux. 
In the far infrared regime Schmitt et al., 1997 use the IRAS fluxes as
we have done for AXJ2254+1146.

\begin{figure}
\psfig{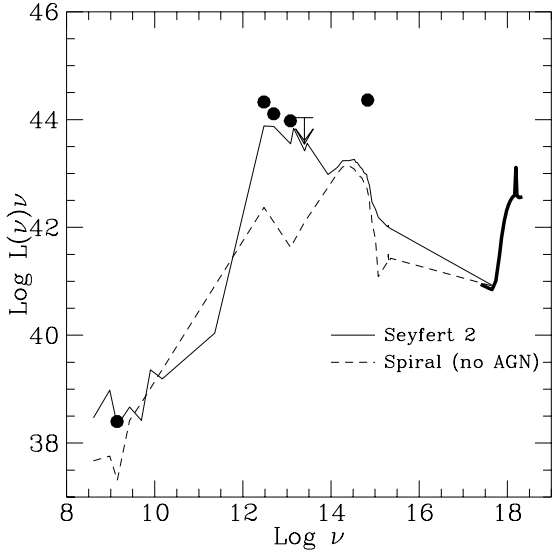}
\caption{SED of AXJ2254+1146: the filled dots represent the photometric
data from radio to optical wavelengths while the solid line represents
the $1 - 10$ keV best fit spectral model.  The arrow represents the
upper limit at 12 $\mu$m.  The other lines represent mean SED of
Seyfert 2 and of non-AGN Spiral galaxies as indicated in the figure.}
\end{figure}

\section {Optical Spectroscopy}


While the paper was in the refereeing process 
an optical spectrum was obtained with VLT/FORS in long slit mode 
as a backup program during a
period of 
poor seeing.
The exposure time was 5 minutes long and the 
slit ($1.6\arcsec$ wide) was set 
perpendicular to the galaxy major axis.
The Grism 150I
was used providing  a resolution
$\Delta\lambda\sim\! 9\AA$ with little sensitivity below $4000\AA$. 
The extracted spectrum (Fig.6) shows 
H$_\beta$, [OIII], [OI], H$_\alpha$, [NII] and [SII]
narrow emission lines 
as well as absorption lines originating in the host galaxy, 
at redshift $z_{\rm opt} = 0.029$.
The weak H$_{\beta}$ line is likely partially obscured by blending with the stellar
H$_{\beta}$ absorption line. We estimate a flux ratio$[\rm{O}{\sc III}]~\lambda
5007/\rm{H_{\beta}} >3$.  
At the above stated spectral resolution, we also estimate 
a FWHM for all the observed emission lines less than 
$\sim 800$ km/s.
Together with the presence of 
a weak [OI]$\lambda 6300$ line, these line features and ratios 
lend support
to the identification of AXJ2254+1146 as a Seyfert 2, as opposed to a LINER or
starburst galaxy.

\begin{figure}
\psfig{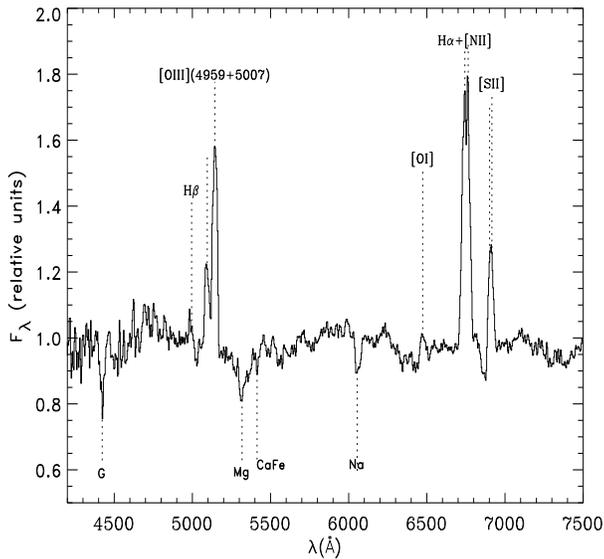}
\caption{Optical spectrum of UGC 12237. See section 5 for details.}
\end{figure}

\section {Summary and Conclusion}

In this paper we have reported the X-ray discovery of a Type 2 AGN -
AXJ2254+1146 - which is part of the ASCA Hard Serendipitous Survey
(Della Ceca et al., 1999a, 2000); this object must be added to the scanty
number of hard X-ray selected type 2 AGN reported in literature (see
e.g. Akiyama et al., 1998b and reference therein).

The X-ray spectrum is best described by a model consisting of an unresolved
Gaussian line at $6.43\pm 0.1$ keV plus the so called ``leaky-absorber"
continua.  The best fit values of the absorbing column density ($N_H$ =
$1.85^{2.24}_{1.47} \times 10^{23}$ cm$^{-2}$), of the line equivalent
width ($0.6^{0.84}_{0.36}$ keV) and of the scattering fraction
($0.7^{1.4}_{0.1} \%$) clearly reveal its Type 2 nature.

A fairly bright ($m_{B^{o}} = 14.26$) spiral galaxy (UGC 12237) is 
present inside the ASCA error circle: this is the most probable 
optical counterpart of AXJ2254+1146.
Subsequent optical spectroscopy  has confirmed that UGC 12237 is 
a Seyfert 2 at z=0.029.

Which are the most important implications of this finding?
As with many discovery of this kind we are probably seeing the ``tip of
the iceberg". As soon as the new generation of X-ray telescopes (e.g.
Chandra, XMM) is fully operative, we predict that it will become
common to investigate and classify X-ray selected objects by using
X-ray data alone (see also Rosati et al., 1995 for a classification based
on the source extent).
AXJ2254+1146 could be indeed one of the brightest examples of a
population of very hard X-ray sources that is showing up at faint
fluxes (cf. Della Ceca et al., 1999a,b). 
The current optical identification status of the ASCA HSS 
prevents us from speculating any further and discussing 
constraints on the synthesis models for the CXB. 
We only note here that in the ASCA HSS sample there are about 8 more
sources with broad band spectral properties (as derived from the
hardness ratios) similar to those of AXJ2254+1146.
One of these 8 sources is NGC 6552, a well known and studied Seyfert 2 galaxy
(Reynolds et al., 1994).
Finally, we recall that Fabian and Iwasawa (1999) have recently pointed
out that if absorbed AGNs are the major contributors to the CXB above 2
keV then about 85\% of the accretion power in the Universe is
absorbed.  In AXJ2254+1146 about 99\% (65\%) of the intrinsic flux in
the 0.5-2.0 (2 -10) keV energy range is absorbed !

\begin{acknowledgements}

We are grateful to the referee, M. Cappi, and to A. Wolter 
for the careful reading
of the manuscript and for useful comments.
V.B.  acknowledges financial support from the {\it Osservatorio 
Astronomico di Brera}.  
This work received partial financial support from the
Italian Ministry for University and Research (MURST)  under grant
Cofin98-02-32.  
This research has made use of the NASA/IPAC
extragalactic database (NED), which is operated by the Jet Propulsion
Laboratory, Caltech, under contract with the National Aeronautics and
Space Administration.  We thank all the members of the ASCA team who
operate  the satellite and maintain the software data analysis and the
archive.

\end{acknowledgements}

\end{document}